# Magneto-optical properties of Co/ZnO multilayer films

**David S Score[1], Marzook Alshammari[1], Qi Feng[1], Harry J Blythe[1], A Mark Fox[1], Gillian A Gehring[1], Zhi-Yong Quan[2], Xiao-Li Li[2], Xiao-Hong Xu[2]**

[1] Department of Physics and Astronomy, Hicks Building, University of Sheffield, Sheffield S3 7RH, UK

[2] School of Chemistry and Materials Science, Shanxi Normal University, Linfen 041004, PR China

Email: d.score@sheffield.ac.uk

**Abstract**. [Co(0.6 nm)/ZnO($x$ nm)]$_{60}$ ($x$= 0.4nm, 3nm) films were deposited on glass substrates then annealed in a vacuum. The magnetisation of the films increased with annealing but not the magnitude of the magneto-optical signals. The dielectric functions Im $\varepsilon_{xy}$ for the films were calculated using the MCD spectra. A Maxwell Garnett theory of a metallic Co/ZnO mixture is presented. The extent to which this explains the MCD spectra taken on the films is discussed.

## 1. Introduction

Magnetically doped semiconductors are one of the most actively studied areas of magnetism. It is not always known the extent to which the observed magnetism is due to ferromagnetic nanoparticles that give an apparent ferromagnetic signal below their blocking temperature. For samples of ZnO doped with cobalt the nanoparticles of interest are metallic cobalt [1]. Magneto-optics is a very powerful technique to investigate this problem as the analysis of the optical response of a composite medium, the Maxwell Garnett theory (MG), is well established and known to be accurate for a wide range of concentrations [2]. It was recently suggested that the Magnetic Circular Dichroism (MCD) of a film of ZnO:Co containing some metallic cobalt was proportional to the magnetisation from the metallic inclusions [3].

In this paper we investigate the effects of nanoparticles on the magnetic and optical properties of Co/ZnO multilayer thin films [4] because this is a system that may be studied both with and without demonstrable cobalt metal nanoparticles. After deposition such samples have most of the Co deposited in the lattice and show magnetic field dependent variable range hopping conductivity [4]. The magnetisation and coercive field of the samples increases dramatically after annealing and nanoparticles of cobalt are detected [5].

The magneto-optics of cobalt nanoparticles have been extensively studied in other oxides, MgO, Al2O3 and $ZrO_2$, all in the visible region 1.5<E<4.5eV [6,7]. We investigate here the extent to which MG theory also works for our system. Deviations from the MG theory will occur if the ZnO lattice has become magnetic hence this analysis will enable us to determine the extent to which this is so. An interesting difference between the previous studies and our investigation of cobalt in ZnO is the relative positions of the oxide band gaps. In ZnO this is at E~3.4eV which is in the range that is studied and hence both the intrinsic magneto-optic effects due to a polarised lattice and the extrinsic effects from the nanoparticles may be studied together. All the other oxides have band gaps further in the ultra violet.

## 2. Samples and experimental methods.

Co/ZnO multilayer films were grown on glass substrates by magnetron sputtering from a Co target and a ZnO target at room temperature. Current-regulated DC power and RF power were applied to the Co target and ZnO target, respectively. The sputtering chamber pressure was reduced to $8.0\times10^{-5}$ Pa before deposition. The sputtering was undergone in an Ar atmosphere at a pressure of 0.8 Pa. After deposition, the samples were annealed in vacuum conditions under a pressure of $2.0\times10^{-4}$ Pa at 400 °C for 120 min. The as-deposited structure was $[Co(0.6\ nm)/ZnO(x\ nm)]_{60}$ ($x$= 0.4, 3), i.e. sixty repetitions of a 0.6nm Co layer followed by either a 0.4nm ZnO layer or a 3nm ZnO layer. These samples are referred to as the 0.4nm and 3nm samples from now on.

Magneto-optical measurements were made at room temperature using a method that enabled the simultaneous measurement of the Faraday rotation and ellipticity [8, 9]. The method uses a pair of polarisers and a photoelastic modulator. A Xenon lamp and monochromator were used to vary the photon energy between 1.6eV and 3.9eV.

## 3. Theoretical modelling of the magneto-optical signals

A magneto-optic signal arises when there is unequal absorption of left and right circularly polarised light. The MCD measures this directly, while the Faraday rotation is present at a given frequency if there is unequal absorption at other frequencies. Both these effects can be ascribed to the off-diagonal terms in the dielectric constant, which for an isotropic medium magnetised in the $\hat{z}$ direction, may be defined as [10],

$$\vec{\varepsilon} = \begin{pmatrix} \varepsilon_{xx} & i\varepsilon_{xy} & 0 \\ -i\varepsilon_{xy} & \varepsilon_{xx} & 0 \\ 0 & 0 & \varepsilon_{zz} \end{pmatrix}. \tag{1}$$

Using the definition above and $\varepsilon_{xy} << \varepsilon_{xx}$ the refractive indices for left and right circularly polarised light are found,

$$n_{\pm} = \sqrt{\varepsilon_{xx} \pm \varepsilon_{xy}} \cong n \pm \frac{\varepsilon_{xy}}{2n} \tag{2}$$

where $n = \sqrt{\varepsilon_{xx}}$ . All the quantities in the above equation will be complex for a film that absorbs light. The MCD, $\eta$, and Faraday rotation, $\theta$, are given for a film of thickness, $L$, at frequency $\omega$ in terms of $n_{\pm}$ by,

$$\eta = \frac{\omega L}{2c}\,\mathrm{Im}\left(n_{+} - n_{-}\right) \text{ and } \theta = \frac{\omega L}{2c}\,\mathrm{Re}\left(n_{+} - n_{-}\right). \tag{3}$$

These equations are solved for $\varepsilon_{xy}$ where we assumed that $n$ for ZnO was real below the band gap [9].

The MG equation for the effective value of $\varepsilon_{xy}$ is given in terms of the dielectric functions for bulk Co, $\varepsilon_{xx}^{Co}, \varepsilon_{xy}^{Co}$, the dielectric function of ZnO, $\varepsilon_{xx}^{ZnO}$, shape factors $L_x$ (=1/3 equal to for spherical particles) and the fraction, $f$, of the sample that is metallic cobalt.

$$\varepsilon_{xy}^{eff} = \frac{f\,\varepsilon_{xy}^{Co}}{A^2}; \qquad A = 1 + L_x(1-f)\left(\frac{\varepsilon_{xx}^{Co}}{\varepsilon_{xx}^{ZnO}} - 1\right) \tag{4}$$

We use the dielectric functions for cobalt, $\varepsilon_{xx}^{Co}, \varepsilon_{xy}^{Co}$, as fitted by Drude theory [10] and $\varepsilon_{xx}^{ZnO}$ from the refractive index of ZnO [11].

In figure 1 we show a plot of the effective dielectric function $\operatorname{Im}\varepsilon_{xy}^{eff}$ for $L_x$ =0.33 and $f$=0.3 for different values of the relaxation time, τ. Previous work had found that the data on Co nanoparticles could be fitted by reducing the value of τ to ~0.2(eV)$^{-1}$. We see that reducing τ reduces and flattens the plot of the dielectric functions but does not change the energy, $\omega_X$ , at which $\operatorname{Im}\varepsilon_{xy}^{eff}(\omega)=0$. This may be shown to occur at $\omega_X^2 = \dfrac{a}{n^2(1-a)+a}\omega_p^2$ where $a= L_x(1-f)$ , $n$ is the refractive index of ZnO and $\omega_p$=9.74eV, the plasma frequency for Co.

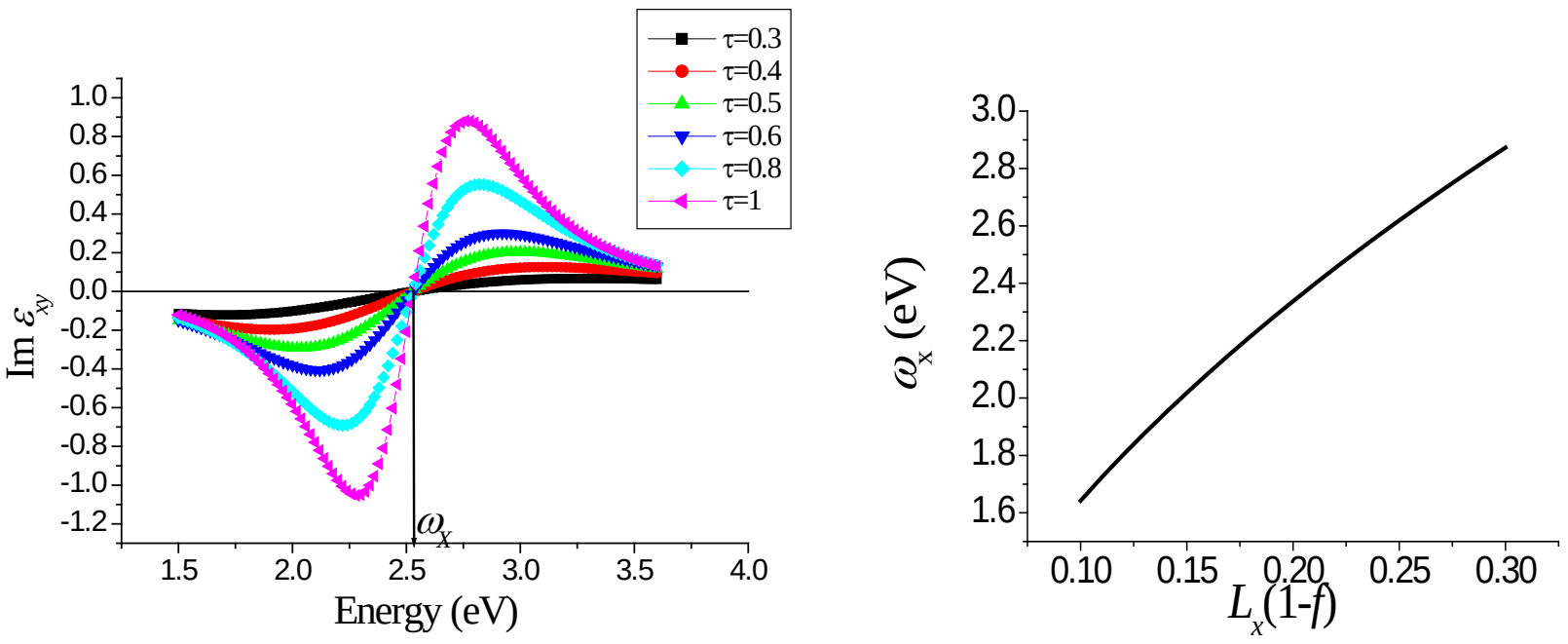

**Figure 1.** Effective dielectric functions $\operatorname{Im}\varepsilon_{xy}^{eff}$ calculated in MG theory for $L_x$=0.33 and $f$=0.3 and different, $\tau$ and the variation of $\omega_X$ with $L_x(1-f)$ .

## 4. Experimental results and discussion

In figure 2, we present room temperature magnetisation loops and $\operatorname{Im}\varepsilon_{xy}$ calculated from MCD spectra before and after annealing. The MG theory, which works well for Co nanoparticles in other

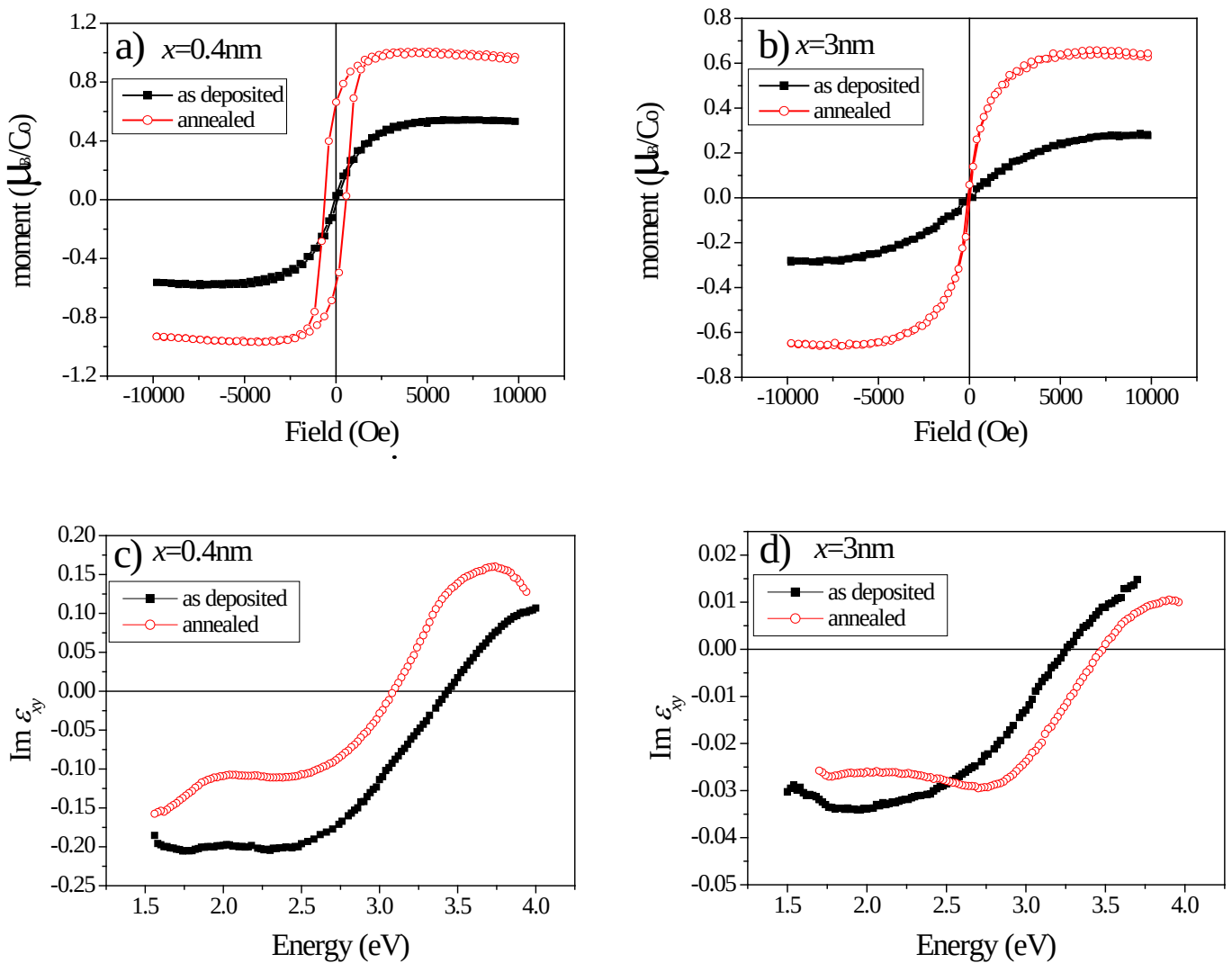

**Figure 2**. Room temperature hysteresis loops of the as-deposited and annealed samples of **(a)** 0.4nm sample and **(b)** 3nm sample. Im $\varepsilon_{xy}$ calculated from magneto-optic spectra of **(c)** 0.4nm sample and **(d)** 3nm sample

oxides [6] is not fitting this data. Theory predicts that the magnitude of Im $\varepsilon_{xy}$ is directly proportional to the fraction of metallic Co. This is not observed. The magnetization increases after the anneal, however there is not a corresponding increase in the value of Im $\varepsilon_{xy}$. We see a substantial signal in both samples before annealing when the magnetic and structural data indicate that metallic Co is absent in measurable concentrations. A clear X-ray signature of metallic Co is present in the 0.4nm sample after the anneal [5] and yet the magnitude of Im $\varepsilon_{xy}$ has decreased for all energies below 3eV. For the 3nm film, the ZnO wurtzite structure is preserved and provided a barrier to large Co cluster formation during the annealing [5]. In both cases annealing resulted in the multilayer structure changing to a granular structure with either inclusions of Co inclusions in ZnO or ZnO inclusions in Co.

Furthermore, the theory predicts a significantly lower value of $\omega_X$ than is observed. If the Co nanoparticles are oblate rather than spherical this will reduce the value of $L_x$ and hence further reduce the predicted $\omega_X$. We note that increasing the fraction of Co nanoparticles also reduces the predicted $\omega_X$. Interestingly, this crossing energy is around the band gap for zinc oxide, $E_g$~3.4eV. A (negative) MCD signal around 3eV from a magnetic ZnO lattice would increase the energy at which Im $\varepsilon_{xy}$=0. The magneto-optic spectra observed here can only be fitted by a model that also includes a contribution from the Co dispersed in the ZnO lattice as was observed previously [12].

## 5. References

[1] Opel M, Nielsen K W, Bauer S, Goennenwein S T B, Cezar J C, Schmeisser D, Simon J, Mader W and Gross R 2008 *Eur. Phys. J.* B **63** 437

Heald S M *et al.* 2009 *Phys Rev* B **79** 075202

[2] Mallet P, Guérin C A and Sentenac A 2005 *Phys Rev* B **72** 014205

[3] Fukuma Y, Asada H, Yamamoto J, Odawara F and Koyanagi T 2008 *Phys Rev* B **78** 104417

[4] Yan S S, Liu J P, Mei L M, Tian Y F, Song H Q, Chen Y X, and Liu G L 2006 *J. Phys Cond. Matt.* **18** 10469 (2006)

[5] Li X L, Quan Z Y, Xu X H, Wu H S and Gehring G A 2008 *IEEE Trans Mag* **44** 2684 and to be published

[6] Clavero C, Armelles G, Margueritat J, Gonzalo J, García del Muro M, Labarta A and Batlle X 2007 *Appl. Phys. Lett.* **90** 182506

Clavero C, Sepúlveda B, Armelles G, Konstantinović Z, García del Muro M, Labarta A and Batlle X 2006 *J. Appl. Phys.* **100** 074320

Clavero C, Cebollada A, Armelles G, Huttel Y, Arbiol J, Peiró F and Cornet A 2005 *Phys Rev* B **72** 02441

[7] One should note that this is the common definition used by those who use Faraday geometry. Those that work in Kerr geometry (including ref [6]) use the alternative definition for the dielectric tensor: $\vec{\vec{\varepsilon}} = \begin{pmatrix} \varepsilon_1 & \varepsilon_2 & 0 \\ -\varepsilon_2 & \varepsilon_1 & 0 \\ 0 & 0 & \varepsilon_3 \end{pmatrix}$ and hence their results are related to ours by $\varepsilon_2' = -\varepsilon_{xy}''$ and $\varepsilon_2'' = \varepsilon_{xy}'$.

[8] Sato K 1981 *Jpn. J. Appl. Phys.* **20** 2403

[9] Van Drent W P and Suzuki T 1997 *J. Magn. Magn. Mater.* **175** 53

[10] Krinchik G S 1964 *J. Appl. Phys.* **35**, 1089

[11] Sun X W and Kwok H S 1999 *J. Appl. Phys.* **86** 408

[12] Neal J R, Behan A J, Ibrahim R M, Blythe H J, Ziese M, Fox A M and Gehring G A 2006 *Phys. Rev. Lett.* **96** 197208